\documentclass[aps,pra,reprint,nofootinbib,superscriptaddress]{revtex4-1}
\usepackage{LatexStyle}
\usepackage[T1]{fontenc}
\usepackage[utf8]{inputenc}
\usepackage{lmodern}
\usepackage{microtype}
\usepackage{amsmath,amssymb,bm,mathtools}
\usepackage{graphicx}
\usepackage{xcolor}
\usepackage{hyperref}
\hypersetup{
    unicode=true,
    pdfusetitle,
    bookmarks=true,
    bookmarksnumbered=true,
    bookmarksopen=false,
    breaklinks=false,
    pdfborder={0 0 0},
    pdfborderstyle={},
    backref=false,
    colorlinks=false,
    pdftitle={Exact Analytical Treatment of Collectively Pumped Dicke Superradiance}
}
\usepackage{physics}
\hypersetup{
  colorlinks=true,
  citecolor=blue,
  linkcolor=black,
  urlcolor=blue
}

\newcommand{\iu}{\mathrm{i}}

\binoppenalty=\maxdimen
\relpenalty=\maxdimen

\begin{document}

\title{Exact Analytical Treatment of Collectively Pumped Dicke Superradiance}

\author{Avishi Poddar}
\email{avishipoddar@g.harvard.edu}
\affiliation{Department of Physics, Harvard University, Cambridge, Massachusetts 02138, USA}

\author{Nico Baßler}
\affiliation{Institute for Applied Physics, TU Darmstadt, Hochschulstra\ss e 4A, D-64289 Darmstadt, Germany}

\author{Julian Lyne}
\affiliation{Department of Physics, Friedrich-Alexander-Universität Erlangen-Nürnberg, Staudtstra\ss e 7, D-91058 Erlangen, Germany}

\author{Susanne F. Yelin}
\affiliation{Department of Physics, Harvard University, Cambridge, Massachusetts 02138, USA}

\author{Helmut Ritsch}
\affiliation{Institute for Theoretical Physics, University of Innsbruck, Technikerstr. 21a, A-6020 Innsbruck, Austria}

\author{Claudiu Genes}
\affiliation{Institute for Applied Physics, TU Darmstadt, Hochschulstra\ss e 4A, D-64289 Darmstadt, Germany}

\author{Raphael Holzinger}
\affiliation{Institute for Theoretical Physics, University of Innsbruck, Technikerstr. 21a, A-6020 Innsbruck, Austria}

\date{\today}

\begin{abstract}
We study the dynamics of an ensemble of $N$ two-level emitters undergoing collective decay at rate $\Gamma$ and collective incoherent pumping at rate $R$, confined to the fully symmetric Dicke ladder. Although the population dynamics are confined to a finite set of rate equations, their quantum-trajectory representation involves infinitely many pump-decay histories. We derive an exact finite continued fraction for the Laplace-transformed Dicke-state populations by resumming the infinitely many quantum trajectories generated by competing pump and decay processes. Our construction provides a physically transparent interpretation of the tridiagonal resolvent and extends to arbitrary permutationally symmetric initial states. In the strong- and weak-pumping regimes, we obtain perturbative transient solutions. We also identify exact relaxation eigenvalues and emission observables at balanced pump and decay strengths. Our solution provides a starting point for extending trajectory-based analytical methods to more general driven-dissipative collective systems.
\end{abstract}
\maketitle

\section{Introduction}

\begin{figure}[ht!]
  \centering
  \includegraphics[width=1\columnwidth]{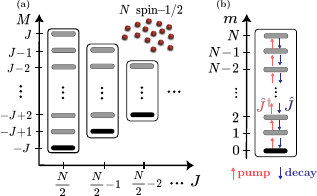}
  \caption{\textbf{Collectively pumped Dicke superradiance.}
 \textbf{(a)} The Hilbert space of $N$ spin-$1/2$ emitters decomposes into Dicke ladders $\ket{J,M}$ of fixed total angular momentum $J$. Vertical transitions $M\rightarrow M\pm1$ preserve $J$, so that each $J$ ladder forms an invariant dynamical subspace, remains decoupled from all other ladders, and contains $2J+1$ states.
 \textbf{(b)} In the fully symmetric ladder $J=N/2$, the dynamics are restricted to the $N+1$ Dicke states $\ket{m}$, labeled by $m=M+N/2=0,\ldots,N$. Collective pumping ($\hat J^\dagger$) and collective dissipation ($\hat J$) induce upward and downward transitions, respectively.}
  \label{fig1}
\end{figure}

Dicke superradiance is a paradigmatic example of cooperative open-system dynamics. An ensemble of identical two-level emitters coupled symmetrically to a common radiation channel can radiate collectively, resulting in strongly enhanced photon-emission intensities relative to independent spontaneous emission~\cite{Dicke1954,RehlerEberly1971,BonifacioLugiato1975,GrossHaroche1982}. Although the dimension of the Hilbert space of $N$ two-level systems grows exponentially as $2^N$, permutational invariance greatly reduces the complexity of the dynamics~\cite{ShammahEtAl2018,HolzingerEtAl2025,HolzingerGenes2024,holzinger2026symbolic}. In general, weak permutational symmetry allows the dynamics of Dicke-state populations to be described within the Dicke triangle, while still permitting transitions between different total-spin sectors $J$, for instance through local spontaneous emission. In contrast, for purely collective processes, the permutational symmetry is strong, and the conservation of total spin decomposes the dynamics into $\lfloor N/2\rfloor+1$ independent Dicke ladders (see \figref{fig1}). This additional structure considerably simplifies the dynamics and enables exact analytical solutions in suitable cases~\cite{Lee1977I,Lee1977II}.

The analytical tractability of collective Dicke dynamics has motivated a long history of exact and asymptotic treatments. Exact solutions of the purely decaying superradiance master equation were obtained for complete and arbitrary initial excitation~\cite{Lee1977I,Lee1977II}, while later work developed useful descriptions for large system sizes~\cite{MalzTrivediCirac2022}. Coherently driven superradiance, or cooperative resonance fluorescence, also admits exact analytical steady-state treatments and correlation functions~\cite{puri1979exact,kilin1980cooperative,drummond1980observables}. More recently, a symbolic quantum-trajectory approach made finite-$N$ time-dependent population dynamics analytically accessible~\cite{HolzingerGenes2024,HolzingerEtAl2025}. Related integrability results for collective spins coupled to polarized Markovian reservoirs include collective raising and lowering channels as a special case and provide exact information about the Liouvillian spectrum and eigenmodes~\cite{RibeiroProsen2019}. Thus, the Dicke ladder provides a rare setting in which genuinely collective many-body open-system dynamics can be treated analytically rather than only through mean-field, cumulant, bosonic, or other approximate methods~\cite{Kubo1962,BonifacioLugiato1978,KramerRitsch2015,BarberenaRey2024}.

Here, we consider Dicke superradiance in the presence of collective incoherent pumping. The steady-state populations follow immediately from detailed balance, but the transient problem is qualitatively richer than pure decay: a given Dicke state can be reached after arbitrarily many alternating pump and decay events. Consequently, an exact quantum-trajectory construction~\cite{DalibardCastinMolmer1992} involves an infinite number of possible jump histories, even though the underlying Dicke ladder contains only finitely many states.

The collectively pumped Dicke model has recently attracted interest in connection with dissipative criticality and metrology. Ref.~\cite{ShimshiShahmoon2024} shows that the sharp finite-size crossover around balanced pumping and decay, $R=\Gamma$, becomes a first-order dissipative transition in the thermodynamic limit, producing enhanced sensitivity to the pump rate. The same work also proposes a direct realization of this model in a cavity QED setup in which an atomic ensemble is coupled to two cavity modes; one mode mediates collective decay on the relevant two-level transition, while the other mode, together with an off-resonant classical laser, couples through an auxiliary atomic level and, after adiabatic elimination, generates an effective collective pumping channel. The resulting reduced dynamics are precisely of the form considered here, with independently controllable collective decay and pump rates.

Despite this progress, a compact exact description of the complete transient population dynamics that explicitly captures the bidirectional pump--decay structure has, to our knowledge, not been presented. While direct numerical diagonalization or integration of the Dicke-state rate equations is straightforward, these approaches do not directly reveal how the infinitely many quantum-jump trajectories are organized or how the continued-fraction structure emerges. Such an analytical representation is useful for deriving controlled approximations in different parameter regimes, and evaluating observables directly from the transient and steady-state dynamics of the Dicke states.

The main result of this work is a finite continued-fraction expression for the Laplace transforms of all Dicke-state populations, together with an equivalent determinant, or continuant, representation that gives the Laplace-transformed populations in closed algebraic form. The continued fraction has a direct trajectory interpretation: each nested level resums repeated upward-and-downward excursions into the next part of the Dicke ladder. We further show that the same structure is precisely the resolvent of a finite-dimensional tridiagonal matrix. The full time-dependent population problem is therefore reduced to a finite spectral problem, yielding an exact sum over relaxation modes. At $R=\Gamma$ the matrix becomes symmetric and its spectrum simplifies to a closed form. At this balanced point, several steady-state observables can likewise be obtained exactly, including the collectively enhanced emission rate, equal-time photon correlations, and the spectral linewidth. Our results provide direct access to transient and steady-state observables and offer a natural starting point for extensions to more general pumped, driven, or multichannel collective systems beyond the symmetric Dicke subspace.

\section{Model}
While $N$ two-level systems generically span a Hilbert space of dimension $2^N$, the collective operators preserve total angular momentum because
$[\hat{\mathbf J}^{2},\hat J]=[\hat{\mathbf J}^{2},\hat J^{\dagger}]=0$.
The dynamics, therefore, separate into Dicke ladders of fixed $J$, as shown in \figref{fig1}(a). For a general ladder, defining $m=M+J$ gives $m=0,\ldots,2J$ and the state-dependent ladder factor
\begin{equation}
h_m^{(J)}\equiv m(2J+1-m).
\end{equation}
In the present work, we consider initial states with $J=N/2$, which confines the evolution to the corresponding ladder containing $N+1$ states, as shown in \figref{fig1}(b). We denote the state with $m$ excitations by $\ket{m}$, with $m=0,\ldots,N$. In this sector,
$h_m^{(J=N/2)}=m(N+1-m)\equiv h_m$, and the collective lowering and raising operators act as
\be\label{eq:collectiveops}
    \hat{J}\ket{m}
    =
    \sqrt{h_m}\ket{m-1},
    \quad
    \hat{J}^{\dagger}\ket{m}
    =
    \sqrt{h_{m+1}}\ket{m+1}.
\ee
All population formulas derived below carry over directly under the replacement $N\rightarrow 2J$. Likewise, a Dicke-diagonal initial state with support on several fixed-$J$ ladders can be propagated ladder by ladder and the resulting contributions appropriately combined.

The states undergo collective decay, described by the action of the collective lowering operator $\hat{J}$, at rate $\Gamma$ and collective pumping, described by the action of the collective raising operator $\hat{J}^\dag$, at rate $R$.
The time evolution of the density operator $\hat{\rho}\pare{t}$ follows the Lindblad master equation, given by
\be\label{eq:mastereq_RGamma}
    \dot{\hat{\rho}} = \!\frac{\Gamma}{2}\left[2\hat{J}\hat{\rho}\hat{J}^\dag \!- \! \left\{\hat{J}^\dag\hat{J},\hat{\rho}\right\}\right] + \frac{R}{2}\left[2\hat{J}^\dag\hat{\rho}\hat{J} \!- \!\left\{\hat{J}\hat{J}^\dag,\hat{\rho}\right\}\right].
\ee

For a permutationally symmetric initial state with populations $p_m(0)$, with $m=0,1,\cdots,N$, the density
operator remains diagonal in the Dicke basis.
Using Eq.~\eqref{eq:collectiveops}, the master equation
\eqref{eq:mastereq_RGamma} then reduces to the population rate
equations
\be\label{eq:rateeq}
\dot{p}_m = \Gamma h_{m+1}p_{m+1} + R h_m p_{m-1}
- \pare{\Gamma h_m + R h_{m+1}}p_m,
\ee
where upward transitions $\ket{m}\to\ket{m+1}$ occur at
rate $Rh_{m+1}$, while downward transitions
$\ket{m}\to\ket{m-1}$ occur at rate $\Gamma h_m$.
The boundary factors $h_0=h_{N+1}=0$ ensure that the
dynamics remain confined to the $N+1$ Dicke states.

Equivalently, the population dynamics can be expressed as
\be
\dot{\mathbf p}(t)=M\mathbf p(t),
\ee
where $\mathbf p(t)=[p_0(t),p_1(t),\ldots,p_N(t)]^T$
and $M$ is the tridiagonal rate matrix
\be
M=\scalebox{0.92}{$\begin{pmatrix}
-Rh_1 & \Gamma h_1 & 0 & \cdots & 0\\
Rh_1 & -(\Gamma h_1+Rh_2) & \Gamma h_2 & \ddots & \vdots\\
0 & Rh_2 & -(\Gamma h_2+Rh_3) & \ddots & 0\\
\vdots & \ddots & \ddots & \ddots & \Gamma h_N\\
0 & \cdots & 0 & Rh_N & -\Gamma h_N
\end{pmatrix}$.}
\ee
Taking the Laplace transform gives the equivalent
resolvent representation
\be\label{eq:resolvent_vector}
\mathbf P(s)=\pare{s\mathbf{1}_{N+1}-M}^{-1}\mathbf p(0),
\ee
where $\mathbf{1}_{N+1}$ is the $(N+1)$-dimensional
identity matrix.

Such tridiagonal matrices are familiar from classical
birth--death processes~\cite{karlin1957classification,karlin1958linear,ismail1990birth,guillemin1999excursion,sasaki2009exactly},
which in special cases admit analytical treatments using
orthogonal polynomials~\cite{koekoek1996askey,koekoek2010hypergeometric}.
While the resolvent formally determines the complete time
evolution, explicit analytical expressions for its matrix
elements are less readily available at general $R$ and
$\Gamma$. In Sec.~\ref{sec:trajectories} we develop a complementary construction
based on quantum trajectories, whose exact resummation
naturally produces a finite continued fraction and makes
the pump--decay processes underlying the transient dynamics
explicit.

\section{Steady state and observables}
\label{sec:steady_state}

Unlike the transient dynamics, the steady-state populations
follow directly from detailed balance. The upward and
downward transition rates satisfy
$Rh_{m+1}p_m^{\rm ss}=\Gamma h_{m+1}p_{m+1}^{\rm ss}$,
so that, with $r\equiv R/\Gamma$, the normalized steady-state
distribution is geometric,
\be\label{eq:steady-state}
p_m^{\rm ss}=\frac{(1-r)r^m}{1-r^{N+1}}.
\ee
At the balanced point $r=1$, it becomes uniform,
$p_m^{\rm ss}=1/(N+1)$.

Due to the particularly simple form of the steady-state
density matrix, steady-state observables can be evaluated
analytically. In particular, the collective emission rate
$I_{\rm ss}=\Gamma\langle\hat J^\dagger\hat J\rangle_{\rm ss}$
is given by
\begin{equation}\label{eq:I_ss}
I_{\rm ss}=\Gamma
\begin{cases}
\dfrac{N(N+2)}{6}, & r=1,\\[8pt]
\dfrac{r\left[N(1-r)(1+r^{N+1})-2r(1-r^N)\right]}
{(1-r^{N+1})(1-r)^2}, & r\neq 1.
\end{cases}
\end{equation}

Likewise, the equal-time second-order correlation function
$g^{(2)}(0)$ can be evaluated analytically (see
App.~\ref{app:observables}). At $R=\Gamma$ it approaches
$g^{(2)}(0)\to6/5$ for $N\gg1$, corresponding to slightly
bunched emission. At the same balanced point, the emission
spectrum follows directly from the quantum-regression theorem.
The uniform steady state leads to the exact first-order
correlation function
$G^{(1)}(\tau)=\langle\hat J^\dagger(\tau)\hat J(0)\rangle_{\rm ss}
=e^{-\Gamma\tau}\langle\hat J^\dagger\hat J\rangle_{\rm ss}$,
whereas for generic $r$ the calculation requires propagating
the relevant coherence sector. In a frame rotating at the
emitter transition frequency, the unnormalized spectrum is
therefore Lorentzian,
\begin{equation}
S(\omega)=2\operatorname{Re}\int_0^\infty d\tau\,
e^{i\omega\tau}G^{(1)}(\tau)
=\frac{N(N+2)}{3}\frac{\Gamma}{\Gamma^2+\omega^2}.
\end{equation}
Its full width at half maximum is $2\Gamma$. Thus, although
the emission intensity scales quadratically with $N$ at
balanced pumping, the spectral linewidth is independent of
the number of emitters.

\section{Quantum trajectories}
\label{sec:trajectories}

Although the steady state follows directly from detailed
balance, the full transient dynamics involve infinitely
many possible jump histories. We therefore unravel the
Lindblad master equation~\eqref{eq:mastereq_RGamma} in terms
of quantum trajectories, generalizing previous analytical
treatments of pure Dicke superradiance~\cite{HolzingerGenes2024,HolzingerEtAl2025,holzinger2026symbolic}.
We show that the competing collective pump and decay events
can be organized into closed excursions and resummed exactly
into a finite continued fraction.

For simplicity, we take the initial state to be the ground state $\ket{0}$. A general solution for arbitrary Dicke-diagonal initial states is presented in App.~\ref{app:trajectories}. In the quantum trajectories picture, the dynamics of such systems can be described by evolution under a non-Hermitian Hamiltonian $\hat{H}_{\mathrm{eff}}$, with quantum jumps occurring at certain points. Evolution under the effective non-Hermitian Hamiltonian
\begin{equation}
    \hat{H}_{\mathrm{eff}} = -\frac{\iu}{2}\big({\Gamma \hat{J}^\dag\hat{J} + R\hat{J}\hat{J}^\dag} \big),
\end{equation}
with $\hat{U}\pare{t} = \mathrm{exp} \big(-\iu\hat{H}_{\mathrm{eff}}t \big)$ can be easily evaluated since it remains diagonal in the Dicke basis. The challenge is dealing with the competing actions of the jump operators $\hat{J}$ and $\hat{J}^\dag$ at rates $\Gamma$ and $R$, respectively.

Consider a trajectory that starts from the ground state $\ket{0}$ and ends in the Dicke state $\ket{k}$. Our derivation generalizes easily to all other initial Dicke states, as shown in App.~\ref{app:anyinitalm}. If $\ell$ denotes the number of downward decay jumps, the trajectory must contain $k+\ell$ upward pump jumps in order to end with $k$ excitations. The total number of jumps is therefore $q=(k+\ell)+\ell=k+2\ell$.
We denote the corresponding jump sequence by $\bm{\alpha}=(\alpha_1,\ldots,\alpha_q)$ and define the jump operators as
\begin{equation}
\hat{L}_{\alpha_j}
=
\begin{cases}
\hat{J}^\dagger, & \alpha_j=+1,\\
\hat{J}, & \alpha_j=-1,
\end{cases}
\end{equation}
where $\alpha_j=+1$ and $\alpha_j=-1$ correspond to upward pump and downward decay jumps, respectively.  For jump times $\bm t_q=(t_1,\ldots,t_q),
0<t_1<\cdots<t_q<t$, the unnormalized quantum trajectory state is then
\begin{widetext}
\begin{equation}
\ket{\psi^{(q)}_{k,\ell,\bm\alpha}(t;\bm t_q)}
=
\sqrt{R^{k+\ell}\Gamma^\ell} \
\hat{U}(t-t_q)
\hat{L}_{\alpha_q}
\hat{U}(t_q-t_{q-1})
\cdots
\hat{L}_{\alpha_2}
\hat{U}(t_2-t_1)
\hat{L}_{\alpha_1}
\hat{U}(t_1)
\ket{0}.
\end{equation}
\end{widetext}

\begin{figure*}[ht]
  \centering
\includegraphics[width=0.8\textwidth]{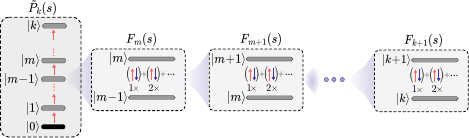}
  \caption{\textbf{Quantum-trajectory resummation.}
    Schematic representation of the exact trajectory construction for the Laplace-transformed population $P_k(s)$, starting from the Dicke ground state $|0\rangle$. The direct upward trajectory $|0\rangle\rightarrow|1\rangle\rightarrow\cdots \rightarrow|k\rangle$ contributes $\widetilde P_k(s)$ [Eq.~\eqref{eq:upward}]. All additional trajectories are generated by inserting closed pump--decay excursions, which are resummed by the continued-fraction factors $F_m(s)$ [Eq.~\eqref{eq:continuedf}]. Each factor $F_m(s)$ accounts for excursions from $|m-1\rangle$ to $|m\rangle$ and back, including nested excursions to higher Dicke states. In particular, $F_{k+1}(s)$ accounts for excursions above the final state $|k\rangle$. Their product yields the exact population $P_k(s)$ [Eq.~\eqref{eq:compact}].}
  \label{fig2}
\end{figure*}

Since infinitely many jump sequences can lead to the same final state, all such histories must be summed. The different possible jump sequences can be thought of in terms of the unidirectional trajectory \(\ket{0}\!\rightarrow\!\ket{1}\!\rightarrow\!\cdots\!\rightarrow\!\ket{k}\) containing \(k\) pump jumps and no decay jumps, with varying numbers of closed excursions between each pair of neighboring Dicke states \(\ket{m-1}\leftrightarrow\ket{m}\), as illustrated in \figref{fig2}. Summing over all possible numbers of such closed excursions yields the full set of allowed trajectories. As detailed in App.~\ref{app:trajectories}, this procedure yields the following continued-fraction form for the Laplace-transformed population of the $k$th Dicke state:
\begin{widetext}\be\label{eq:Pk_anyN}
    P_k\pare{s}= R^k h_1 \ ... \ h_k U_0^2\pare{s}U_1^2\pare{s} \ ... \ U_k^2\pare{s} \frac{1}{1-\frac{C_1\pare{s}}{1-\frac{C_2\pare{s}}{1-...\frac{C_{N-1}\pare{s}}{1-C_N\pare{s}}}}}
    \frac{1}{1-\frac{C_{2}\pare{s}}{1-  ... \frac{C_{N-1}\pare{s}}{1-C_N\pare{s}}}} \ \ ... \ \ \frac{1}{1-\frac{C_{k+1}\pare{s}}{1-...\frac{C_{N-1}\pare{s}}{1-C_N\pare{s}}}},
\ee\end{widetext}
where the no-jump population propagator
\begin{equation}
U_m^2(s) =\frac{1}{\pare{s+\Gamma h_m+R h_{m+1}}},
\end{equation}
is the Laplace transform of $u_m\pare{t}\equiv|\langle m | \hat U(t) |m\rangle|^2$ and
\begin{equation}
    C_m\pare{s}\equiv R\Gamma h_m^2 U_{m-1}^2\pare{s}U_m^2\pare{s},
    \qquad m=1,\ldots,N,
\end{equation}
with the boundary convention $C_{N+1}(s)\equiv0$. Each factor $C_m \propto R\Gamma$ introduces one pump--decay pair, representing the closed loop
\begin{equation}
    |m-1\rangle
    \xrightarrow{R}
    |m\rangle
    \xrightarrow{\Gamma}
    |m-1\rangle. \nonumber
\end{equation}

\subsection{Continued-fraction interpretation}
We can rewrite \eqnref{eq:Pk_anyN} in a physically transparent form,
\begin{equation} \label{eq:compact}
    P_k(s)
    =
    \widetilde{P}_k(s)
    F_1(s)F_2(s)\cdots F_{k+1}(s),
\end{equation}
where
\begin{equation} \label{eq:upward}
    \widetilde{P}_k(s)
    =
    R^k h_1\cdots h_k
    \prod_{j=0}^{k}U_j^2(s)
\end{equation}
is the contribution stemming from the trajectory $|0\rangle\rightarrow |1\rangle\rightarrow\cdots\rightarrow |k\rangle$, 
containing exactly $k$ upward jumps and 
\begin{equation} \label{eq:continuedf}
    F_m(s)
    =
    \frac{1}{1-C_m(s)F_{m+1}(s)} =
    \sum_{n=0}^{\infty}
    \left(C_mF_{m+1}\right)^n,
\end{equation}
represent the continued-fraction factors accounting for all possible closed pump-decay excursions, with $F_{N+1}\pare{s}=1$ reflecting the finite nesting depth of the Dicke ladder.

\subsection{Time-domain solution}

The continued-fraction solution in Eq.~\eqref{eq:Pk_anyN} can equivalently be written in terms of continuants, i.e. determinants of tridiagonal matrices. In particular,
\be\label{eq:Pk_continuant}
P_k(s)
=
R^k h_1\cdots h_k
U_0^2(s)\cdots U_k^2(s)
\frac{D_{k+2}(s)}{D_1(s)},
\ee
where
\be
D_m(s)=\det\! \scalebox{0.92}{$\begin{pmatrix}
1 & -C_m(s) & 0 & \cdots & 0\\
-1 & 1 & -C_{m+1}(s) & \ddots & \vdots\\
0 & -1 & 1 & \ddots & 0\\
\vdots & \ddots & \ddots & 1 & -C_N(s)\\
0 & \cdots & 0 & -1 & 1
\end{pmatrix}$},
\ee
with $D_{N+1}(s)=D_{N+2}(s)=1$. These continuants satisfy the recursion
\begin{equation}
    D_m(s)=D_{m+1}(s)-C_m(s)D_{m+2}(s),
\end{equation}
so that $D_m/D_{m+1}=1-C_m D_{m+2}/D_{m+1}$ recursively reproduces the nested continued fraction in Eq.~\eqref{eq:Pk_anyN}. The determinant representation therefore provides a compact algebraic form of the exact trajectory resummation. 

The possible poles of $P_k(s)$ are given by the eigenvalues $\{\lambda_0,\ldots,\lambda_N\}$ of the tridiagonal rate matrix introduced above. For
$R,\Gamma>0$, this matrix is similar to a real symmetric Jacobi matrix and consequently has real, nondegenerate eigenvalues. The inverse Laplace transform can therefore be evaluated exactly by contour integration,
\be
p_k(t)
=
\frac{1}{2\pi i}
\oint_{\mathcal C}
ds\,e^{st}P_k(s),
\ee
where the positively oriented contour $\mathcal C$
encloses all poles $s=\lambda_j$ in the complex plane. Thus, although the quantum-trajectory representation involves infinitely many jump histories, their exact resummation produces at most $N+1$ distinct relaxation modes.

For generic $R\neq\Gamma$, the eigenvalues do not admit a simple closed analytical expression since this is equivalent to solving a Heun differential equation with four singularities. However, in certain limiting regimes, the loop expansion yields particularly transparent approximations that allow us to analytically evaluate the populations and the time-dependent collective emission rate.

\subsubsection{Balanced case, $R=\Gamma$}

At the balanced point $R=\Gamma$, the continued-fraction
representation in Laplace space remains exact, although
obtaining an explicit time-dependent solution by direct
Laplace inversion is less straightforward. Alternatively,
the tridiagonal population rate matrix yields a three-term
recurrence relation for its eigenmodes, which can be solved
analytically in terms of Hahn polynomials. This yields
the exact eigenvalue spectrum
$\lambda_j=-\Gamma j(j+1)$, with $j=0,\ldots,N$,
and hence a spectral representation of the full
time evolution.

\subsubsection{Weak pumping, $R\ll\Gamma$}

In the weak-pumping regime we retain all contributions consistently to first order in $R$. Starting from the ground state, reaching $|k\rangle$ requires at least $k$ pump jumps; it is therefore sufficient to determine $p_0(t)$ and $p_1(t)$.

For the first excited state, the leading contribution is the direct trajectory
$|0\rangle\rightarrow|1\rangle$. From Eq.~\eqref{eq:upward},
\begin{equation}
P_1(s)
=
Rh_1U_0^2(s)U_1^2(s)
+\mathcal O\pare{\frac{R^2}{\Gamma^2}},
\end{equation}
since every additional loop introduces another factor of $R$ and hence
\begin{equation}
P_1(s)
=\frac{Rh_1}{s(s+\Gamma h_1)}
+\mathcal O\pare{\frac{R^2}{\Gamma^2}}.
\end{equation}
Inverse Laplace transformation gives
\begin{equation}
p_1(t)
=\frac{R}{\Gamma}
\left(1-e^{-\Gamma h_1t}\right)
+\mathcal O\pare{\frac{R^2}{\Gamma^2}}.
\end{equation}

For the ground-state population, the first closed excursion $|0\rangle\rightarrow|1\rangle\rightarrow|0\rangle$ already contributes at order $R$. Expanding the exact trajectory expression,
\begin{equation}
P_0(s)
=U_0^2(s)\left[1+C_1(s)+\mathcal O\pare{\frac{R^2}{\Gamma^2}}\right],
\end{equation}
with
\begin{align}
U_0^2(s)
&\simeq
\frac{1}{s}
-\frac{Rh_1}{s^2},\
\ C_1(s) \simeq
\frac{R\Gamma h_1^2}
{s(s+\Gamma h_1)},
\end{align}
yields
\begin{equation}
P_0(s)
= \frac{1}{s}
-\frac{Rh_1}{s(s+\Gamma h_1)}
+\mathcal O\pare{\frac{R^2}{\Gamma^2}}.
\end{equation}
Therefore,
\begin{equation}
p_0(t)
=
1-
\frac{R}{\Gamma}
\left(1-e^{-\Gamma h_1t}\right)
+\mathcal O[(R/\Gamma)^2].
\end{equation}

Since $h_1=N$, the collective emission rate to the same order becomes
\begin{align}
I(t)
&\simeq
\Gamma h_1p_1(t)
\simeq
RN\left(1-e^{-\Gamma Nt}\right).
\label{eq:weak_emission}
\end{align}
Thus, for $R\ll\Gamma$, the dynamics are confined to the bottom of the Dicke ladder: rare pump events populate $|1\rangle$, which decays collectively on the timescale $(\Gamma N)^{-1}$, while higher Dicke states only enter at second order in $R/\Gamma$.

\subsubsection{Strong pumping, $R \gg \Gamma$}

In the strong-pumping regime, we expand in the small parameter $\Gamma/{R}\ll1$. In contrast to weak pumping, the zeroth-order dynamics already explore the entire Dicke ladder: for $\Gamma=0$, the system follows the monotonic trajectory $|0\rangle\rightarrow|1\rangle\rightarrow\cdots\rightarrow|N\rangle.$ Each additional pump--decay loop contains one factor of $\Gamma$ and therefore produces a relative correction of order $\Gamma/R$.

To zeroth order in $\Gamma/R$, the exact trajectory solution reduces to
\begin{equation}
P_k^{(0)}(s)=
R^k\prod_{m=1}^{k}h_m
\prod_{j=1}^{k+1}
\frac{1}{s+Rh_j},
\quad 0\leq k<N,
\end{equation}
where we used
\begin{equation}
U_m^2(s)=
\frac{1}{s+Rh_{m+1}}
+\mathcal O\pare{\frac{\Gamma}{R^2}}.
\end{equation}
This is essentially the inverse of the purely decaying Dicke superradiance problem.
Since the rates $Rh_j$ are degenerate because $h_j=h_{N+1-j}$, the pole structure of $P_k^{(0)}(s)$ differs based on the choice of $k$. For even (odd) $N$, the poles are simple for $k<N/2 \pare{(N+1)/2}$, while the solution also contains double poles otherwise.

Since the collective emission rate contains an explicit factor of $\Gamma$, its leading contribution only requires the zeroth-order populations,
\begin{equation}
I(t)=
\Gamma\sum_{k=0}^{N}h_kp_k^{(0)}(t)
+\mathcal O(\Gamma^2/R),
\end{equation}
where $p_k^{(0)}(t)$ is the inverse Laplace transform of $P_k^{(0)}(s)$, evaluated explicitly in App.~\ref{app:onlyRsol}.

\section{Conclusion and Outlook}

We have presented an exact analytical solution for the Dicke-state populations of an ensemble of $N$ two-level emitters undergoing collective pumping and collective decay. By organizing quantum-jump histories into nested closed pump-decay excursions, we obtain an exact finite resummation of infinitely many trajectories in the form of a continued fraction. Beyond providing an exact solution, this construction offers a systematic way to enumerate and classify quantum-jump histories according to the number and ordering of pump and decay events, while explicitly accounting for the transition rates and waiting times between successive jumps. This trajectory-resolved perspective provides insight into the pathways and characteristic timescales underlying the collective dynamics. 

The trajectory formulation also provides a natural hierarchy of approximations. In the weak pumping regime, $R\ll\Gamma$, the leading dynamics are confined to the two lowest Dicke states, yielding a simple analytical transient emission rate. In the opposite limit of strong pumping, $\Gamma\ll R$, the zeroth-order dynamics reduce to an inverted version of pure Dicke superradiance.
Precisely at the balanced point $R=\Gamma$, 
the spectral problem simplifies independently, yielding the exact relaxation eigenvalues $\lambda_j=-\Gamma j(j+1)$. Moreover, the uniform steady state in this case leads to closed expressions for the emission intensity, photon correlations, and spectral linewidth.

More broadly, the present results illustrate how exact structure can emerge in driven-dissipative many-body systems despite an apparently unbounded hierarchy of quantum-jump histories. This perspective is complementary to other recent exact approaches to interacting open quantum systems, including integrable collective-spin Liouvillians~\cite{RibeiroProsen2019}, hidden-time-reversal constructions~\cite{RobertsLingenfelterClerk2021}, exact solutions of infinite-range dissipative spin models~\cite{RobertsClerk2023Ising}, and exactly solvable driven-dissipative spin chains~\cite{YaoEtAl2025}. These results suggest that analytically tractable nonequilibrium quantum dynamics extend well beyond the simplest collective models. Analyzing and understanding the properties of continued fractions could provide a promising avenue to study more complex driven-dissipative many-body systems~\cite{hartmann2026exactdampingbasiscontinuedfractions}.

An interesting direction for future work is to determine how far the trajectory-resummation framework itself can be generalized. Extensions to additional dissipative channels, coherent driving, individual pumping, or coupled collective subsystems naturally lead to higher-dimensional trajectory cascades but may retain recursive or continued-fraction structures. Particularly interesting candidates for such generalizations are Rydberg-atom ensembles and multilevel atomic systems, where multiple excitation and decay pathways give rise to more intricate networks of quantum-jump trajectories. Identifying recursive structures within these networks could enable analogous analytical treatments and provide insight into the relative importance of different decay channels and dynamical timescales. Similar ideas may also be applicable to bosonic systems. In number-state representations, bosonic gain and loss processes again generate ladder-like quantum-jump trajectories, now with occupation-dependent bosonic enhancement factors and, in general, an infinite ladder. Truncation, generating-function, or continued-fraction techniques may therefore provide bosonic analogues of the construction developed here. This possibility is particularly intriguing in view of exact results already obtained for interacting driven-dissipative bosonic models~\cite{RobertsClerk2023Bosonic,RobertsLingenfelterClerk2021}. Establishing when such trajectory cascades admit exact resummation could provide a systematic route toward analytical solutions of a broader class of interacting open quantum systems.

\textbf{Acknowledgments} S.F.Y. and A.P. would like to acknowledge the NSF via the CUA PFC (PHY-2317134), via PHY-2608210, and the ARO through the QEMISD MURI project funded by ARO-W911NF-24-2-0195 P00001. R.H. and H.R. acknowledge the Austrian Science Fund (FWF) [10.55776/COE1]. N.B. and C.G. acknowledge financial support from the Max Planck Society and the Deutsche Forschungsgemeinschaft (DFG, German Research Foundation) – Project-ID 429529648 – TRR 306 QuCoLiMa
(“Quantum Cooperativity of Light and Matter”).

\newpage
\appendix
\onecolumngrid

\section{Exact steady-state observables}
\label{app:observables}

As mentioned in the main text, the steady-state populations $p_m^{\text{ss}}$ of the collectively pumped Dicke system take a particularly simple form, given by
\be
    p_m^{\text{ss}} = \begin{cases}
        \frac{1}{N+1} &, r=1,
        \\
        \frac{\pare{1-r}r^m}{1-r^{N+1}} &, r\neq 1.
    \end{cases}
\ee

Using \eqnref{eq:collectiveops}, the photon emission rate through the collective decay channel is simply 
\begin{equation}
    I_{\rm ss}=
    \Gamma\langle \hat J^\dagger\hat J\rangle_{\rm ss}
    = \begin{cases}
        \Gamma \frac{N(N+2)}{6}&, r=1, \\
        \frac{\Gamma r\pare{N\pare{1-r}\pare{1+r^{N+1}}-2r\pare{1-r^N}}}{\pare{1-r^{N+1}}\pare{1-r}^2} &, r\neq1,
    \end{cases}
\end{equation}
as mentioned in \eqnref{eq:I_ss}. Note that for $r=1$, the collective photon emission rate therefore scales as $N^2$. 

The zero-delay normalized second-order correlation function of the emitted field can be similarly computed to obtain
\begin{equation}
    g^{(2)}(0) = \frac{
    \langle
    \hat J^\dagger\hat J^\dagger
    \hat J\hat J
    \rangle_{\rm ss}
    }{
    \langle\hat J^\dagger\hat J\rangle_{\rm ss}^2
    }
    = \begin{cases}
        \frac{6}{5}
        \frac{(N-1)(N+3)}{N(N+2)} \rightarrow \frac{6}{5} \ \mathrm{for} \ N\gg1&, r=1, \\
        \frac{2 (1 - r^{N+1}) (6 r (1 + r) (1 - r^N) + N^2 (1 - r)^2 (1 - r^{N+1})  - N (1 - r) (1 + r (5 + r^N (5 + r))))}{(2 r (1 - r^N) -  N (1 - r) (1 + r^{N+1}))^2} &, r\neq1.
    \end{cases}
\end{equation}

Finally, for the balanced case $r=1$, the spectral linewidth is computable directly from the steady-state solution. The steady-state emission spectrum is determined by the first-order two-time correlation function
\begin{equation}
    G^{(1)}(\tau)
    =
    \langle
    \hat J^\dagger(\tau)\hat J(0)
    \rangle_{\rm ss} =  \operatorname{Tr}
    \left[
    \hat J^\dagger
    e^{\mathcal L\tau}
    \left(
    \hat J\rho_{\rm ss}
    \right)
    \right],
\end{equation}
using the quantum regression theorem. 
It is therefore sufficient to consider the nearest-off-diagonal coherence sector. Writing
\begin{equation}
    e^{\mathcal L\tau}
    \left(
    \hat J\rho_{\rm ss}
    \right)
    =
    \sum_{m=1}^{N}
    x_m(\tau)\ket{m-1}\bra{m},
\end{equation}
the master equation, applied to each coherence $|m-1\rangle \langle m|$, yields
\begin{align}
    \dot{x}_m
    ={}&
    R\sqrt{h_{m-1}h_m}\,x_{m-1}
    +
    \Gamma\sqrt{h_mh_{m+1}}\,x_{m+1}
    -
    \frac{\Gamma}{2}
    \left(
    h_{m-1}+h_m
    \right)x_m
     -
    \frac{R}{2}
    \left(
    h_m+h_{m+1}
    \right)x_m.
    \label{eq:coherence_eom_RGamma}
\end{align}
Using
$r x_{m-1}=x_m\sqrt{h_{m-1}/h_m}$,
$x_{m+1}=rx_m\sqrt{h_{m+1}/h_m}$,
with the boundary convention $x_0=x_{N+1}=0$, 
\begin{align}
    \dot{x}_m
    &=
    \Gamma x_m
    \left[
    h_{m-1}+rh_{m+1}
    -
    \frac{1}{2}
    \left(
    h_{m-1}+h_m
    \right)
    -
    \frac{r}{2}
    \left(
    h_m+h_{m+1}
    \right)
    \right].
\end{align}
In general, solving this equation requires the full transient solution. However, at $r=1$, the right-hand side of this equation becomes independent of $m$. In particular, at this special point, 
\begin{align}
    \dot{x}_m =
    \frac{\Gamma}{2}
    \left(
    h_{m-1}+h_{m+1}-2h_m
    \right)x_m = -\Gamma x_m,
\end{align}

Hence,
\begin{equation}
    e^{\mathcal L\tau}
    \left(
    \hat J\rho_{\rm ss}
    \right)
    =
    e^{-\Gamma\tau}
    \hat J\rho_{\rm ss}.
    \label{eq:eigenoperator_RGamma}
\end{equation}
The two-time correlation function therefore reduces exactly to
\begin{equation}
    G^{(1)}(\tau)
    =
    e^{-\Gamma\tau}
    \langle
    \hat J^\dagger\hat J
    \rangle_{\rm ss}
    =
    \frac{N(N+2)}{6}
    e^{-\Gamma\tau}.
    \label{eq:G1_RGamma}
\end{equation}

In a frame rotating at the emitter transition frequency, the unnormalized Lorentzian spectrum is
\begin{equation}
    S(\omega)
    =
    2\,\operatorname{Re}
    \int_0^\infty
    d\tau\,
    e^{i\omega\tau}
    G^{(1)}(\tau) = \frac{N(N+2)}{3}
    \frac{\Gamma}{\Gamma^2+\omega^2}.
\end{equation}
The full width at half maximum is therefore $2\Gamma$. Notably, while the emitted intensity scales quadratically with $N$, the spectral linewidth at the symmetric point is independent of the emitter number. 

\section{Detailed derivation of the continued fraction expression $P_k(s)$}\label{app:trajectories}

Applying the idea presented in the main text, we account for all possible trajectories by taking the bare sequence 
\be
    \ket{0}\to\ket{1}\to\cdots\to\ket{k},
\ee
and inserting in between each jump, all possible closed pump-decay excursions. We work in Laplace space; this allows us to compute the population of a state expressed as a series of nested convolutions by using the property \be\label{eq:LaplaceTransform_1}
    \mathcal{L}\left[f_1\pare{t}*f_2\pare{t}\right] = \mathcal{L}\left[f_1\pare{t}\right]\mathcal{L}\left[f_2\pare{t}\right] \equiv F_1\pare{s}F_2\pare{s},
\ee
where $F_j\pare{s}$ is the Laplace transform of $f_j\pare{t}$.

To understand how to deal with counting all possible trajectories, we can start by just considering the two topmost states of the ladder, $\ket{N-1}$ and $\ket{N}$. For one sequence $\ket{N-1}\to\ket{N}\to\ket{N-1}$, we get, in Laplace space,
\be
    C_N\pare{s} = R \Gamma h_N^2 U_{N-1}^2\pare{s} U_{N}^2\pare{s},
\ee
where $R h_N$ comes from the transition $\ket{N-1}\to\ket{N}$, $U_{N}^2\pare{s}$ from no-jump time evolution in $\ket{N}$, $\Gamma h_N$ from the decay $\ket{N}\to\ket{N-1}$, and finally $U_{N-1}^2\pare{s}$ from no-jump evolution in $\ket{N-1}$. Summing over the infinite possible iterations of this sequence yields
\be\label{eq:nm1loop}
    \sum\limits_{n=0}^\infty \pare{C_N\pare{s}}^n = \frac{1}{1-C_N\pare{s}}.
\ee

Now, if we also consider the $\ket{N-2}$ state, in one sequence, we account for the $\ket{N-2}\to\ket{N-1}$ transition which yields a factor of $R h_{N-1}$. Time evolution in $\ket{N-1}$ gives $U_{N-1}^2\pare{s}$. Then we insert the infinite closed loops on the state $\ket{N-1}$, computed in \eqnref{eq:nm1loop}, and then the jump down from $\ket{N-1}$ to $\ket{N-2}$, which yields a factor of $\Gamma h_{N-1}$. Finally, time evolution in $\ket{N-2}$ gives $U_{N-2}^2\pare{s}$. Once again, if we summed over all (infinite) possible such loops, we get,
\be
    \sum\limits_{n=0}^\infty \pare{\frac{R\Gamma h_{N-1}^2 U_{N-2}^2\pare{s} U_{N-1}^2\pare{s}}{1-C_N\pare{s}}}^n = \sum\limits_{n=0}^\infty \pare{\frac{C_{N-1}\pare{s}}{1-C_N\pare{s}}}^n = \frac{1}{1-\frac{C_{N-1}\pare{s}}{1-C_N\pare{s}}}.
\ee

Continuing this down to the state $\ket{0}$ then gives us the Laplace-space population of the zero-excitation state:
\be\begin{split}\label{eq:P0_anyN}
    P_0\pare{s}
    \!&= \!U_0^2\sum\limits_{\beta_0=0}^\infty \!
    \pare{\!\!R\Gamma h_1^2 U_0^2\pare{s} U_1^2 \!
    \sum\limits_{\beta_1=0}^\infty
    \pare{ \!\!R\Gamma h_2^2 U_1^2U_2^2 \!\sum\limits_{\beta_2=0}^\infty \! \! \pare{\!\!...\! \!\pare{R\Gamma h_{N-1}^2 U_{N-2}^2
    U_{N-1}^2 \!\!\!\sum\limits_{\beta_{N-1}=0}^\infty \!\!\pare{R\Gamma h_N^2 U_{N-1}^2U_N^2}^{\beta_{N-1}\!}\! }\!\!...\!\!\!}^{\beta_2}\!}^{\beta_1}\!\!}^{\beta_0} \\
    &= \frac{U_0^2\pare{s}}{1-\frac{C_1\pare{s}}{1-\frac{C_2\pare{s}}{1-...\frac{C_{N-1}\pare{s}}{1-C_N\pare{s}}}}},
\end{split}\ee
where $C_k\pare{s} \equiv R\Gamma h_k^2 U_{k-1}^2\pare{s}U_k^2\pare{s}$. We note that in this derivation, we used the formula for the infinite sum, which requires $\abs{C_k\pare{s}}<1$. Let us prove that this is true.
\be\begin{split}
    C_k\pare{s} &\equiv R\Gamma h_k^2 U_{k-1}^2\pare{s}U_k^2\pare{s} = \frac{R\Gamma h_k^2}{\pare{s+\Gamma h_{k-1}+R h_k}\pare{s+\Gamma h_k+R h_{k+1}}}
    \\
    &= \frac{R\Gamma h_k^2}{\spare{s^2 +\pare{\Gamma h_{k-1}+R h_k+\Gamma h_k+R h_{k+1}}s + \Gamma^2 h_{k-1}h_k + R\Gamma h_{k-1}h_{k+1} + R^2 h_k h_{k+1} }+ R\Gamma h_k^2}
\end{split}.
\ee
Since each term inside $\spare{...}$ in the denominator is non-negative, it follows that $\abs{C_k\pare{s}}<1$.

To obtain, in general, the population of the $k$th state, we can apply the same idea, but now starting with the $P_0\pare{s}$ solution. Defining 
\be
    Q_m = R h_m U_m^2 \sum\limits_{\beta_m=0}^\infty \pare{...\pare{R\Gamma h_{N-1}^2 U_{N-2}^2 U_{N-1}^2 \sum\limits_{\beta_{N-1}=0}^\infty \pare{R\Gamma h_N^2 U_{N-1}^2U_N^2}^{\beta_{N-1}} }...}^{\beta_m} = \frac{Rh_m U_m^2\pare{s}}{1-\frac{C_{m+1}\pare{s}}{1-...\frac{C_{N-1}\pare{s}}{1-C_N\pare{s}}}},
\ee
the population of the $k$th state is given by 
\be\begin{split}\label{eq:Pk_anyN_app}
    P_k\pare{s}
    &= Q_1\pare{s}Q_2\pare{s}...Q_k\pare{s} P_0\pare{s}\\
    &= R^k h_1...h_k U_0^2\pare{s}U_1^2\pare{s}...U_k^2\pare{s} \frac{1}{1-\frac{C_1\pare{s}}{1-\frac{C_2\pare{s}}{1-...\frac{C_{N-1}\pare{s}}{1-C_N\pare{s}}}}} \frac{1}{1-\frac{C_{2}\pare{s}}{1-...\frac{C_{N-1}\pare{s}}{1-C_N\pare{s}}}}...\frac{1}{1-\frac{C_{k+1}\pare{s}}{1-...\frac{C_{N-1}\pare{s}}{1-C_N\pare{s}}}},
\end{split}\ee
which is exactly \eqnref{eq:Pk_anyN}. 

\subsection{A simple example: $N=2$}\label{app:n2}
As a simple example, we present explicitly this method for $N=2$.
If we start in $\ket{0}$, now we must account for all infinite possible sequences of jumps, with the constraint that no more than two jumps up or down can occur consecutively since $\hat{J}\ket{0}=0$ and $\hat{J}^\dag\ket{2}=0$. The no-jump propagator described by evolution under the non-Hermitian H
Hamiltonian is given by $\hat{U}\pare{t} = e^{-Rt}\ket{0}\bra{0}+e^{-\pare{\Gamma+R}t}\ket{1}\bra{1} +e^{-\Gamma t}\ket{2}\bra{2}$. A given closed sequence for $m=0$ involves a jump up to $m=1$, a closed loop of infinite possible jumps between $m=1$ and $m=2$, then a jump back to $m=0$ and non-Hermitian no-jump evolution between each jump event. Summing over all possible multiples of the jumps yields the population of the $\ket{0}$ state:
\be\begin{split}\label{eq:P0_N2}
    p_0\pare{t} = \sum\limits_{\beta=0}^\infty
    \pare{4R\Gamma}^\beta
    \int\limits_{0<t_1<...<t_{2\beta}<t} dt_1 ... dt_{2\beta} 
    e^{-2R\pare{t-t_{2\beta}}}
    X\pare{t_{2\beta-1},t_{2\beta}}
    ...
    X\pare{t_3,t_4}
    e^{-2R\pare{t_3-t_2}}
    X\pare{t_1,t_2}
    e^{-2Rt_1}
    ,
\end{split}\ee
where \be X\pare{t_i,t_j} = \sum\limits_{\alpha=0}^\infty          \pare{4R\Gamma}^\alpha 
    \!\!\!\!\!\!\!\!\!\int\limits_{t_i<\tilde{t}_1<...<\tilde{t}_{2\alpha}<t_j} \!\!\!\!\!\! \!\!\! \!\!\!\!\! d\tilde{t}_1... d\tilde{t}_{2\alpha} 
    e^{-2\pare{\Gamma+R}\pare{t_j-\tilde{t}_{2\alpha}}}
    e^{-2\Gamma\pare{\tilde{t}_{2\alpha}-\tilde{t}_{2\alpha-1}}}
    ...
    e^{-2\pare{\Gamma+R}\pare{\tilde{t}_3-\tilde{t}_2}}
    e^{-2\Gamma\pare{\tilde{t}_2-\tilde{t}_1}}
    e^{-2\pare{\Gamma+R}\pare{\tilde{t}_1-t_i}}.\ee 
    
Let us first simplify $X\pare{t_i,t_j}$ since this should be quite similar to the $N=1$ solution. To put it into the standard nested convolution form, we do a change of variables $\tau_k = \tilde{t}_k-t_i$. We also define $t_{ij}\equiv t_j-t_i$.
\be\begin{split}
    X\pare{t_i,t_j} &=  \sum\limits_{\alpha=0}^\infty          \pare{4R\Gamma}^\alpha  \!\!\!\!\!\!\!\!\!\!\!\!\!
    \int\limits_{0<\tau_1<...<\tau_{2\alpha}<t_j-t_i} \!\!\!\!\!\!\!\!\!\!\!\!\! d\tau_1... d\tau_{2\alpha} 
    e^{-2\pare{\Gamma+R}\pare{t_j-t_i-\tau_{2\alpha}}}
    e^{-2\Gamma\pare{\tau_{2\alpha}-\tau_{2\alpha-1}}}
    ...
    e^{-2\pare{\Gamma+R}\pare{\tau_3-\tau_2}}
    e^{-2\Gamma\pare{\tau_2-\tau_1}}
    e^{-2\pare{\Gamma+R}\tau_1}
    \\
    &= \sum\limits_{\alpha=0}^\infty \pare{4R\Gamma}^\alpha 
    e^{-2\pare{\Gamma+R}t_{ij}}*\left[ \pare{e^{-2\Gamma t_{ij}}*e^{-2\pare{\Gamma+R}t_{ij}}} * \pare{e^{-2\Gamma t_{ij}}*e^{-2\pare{\Gamma+R}t_{ij}}} * .... \pare{\text{$\alpha$ times}}\right].
\end{split}\ee
Using the commutative property of convolutions,
\be\begin{split}\label{eq:X}
    X\pare{t_i,t_j} &= \sum\limits_{\alpha=0}^\infty \frac{\pare{4R\Gamma}^\alpha }{\alpha!\pare{\alpha-1}!}
    \pare{e^{-2\pare{\Gamma+R}t_{ij}} t_{ij}^{\alpha}}
    *\pare{e^{-2\Gamma t_{ij}}t_{ij}^{\alpha-1}} \\ 
    &= \frac{\sqrt{\pi Rt_{ij}}}{\sqrt{2}}e^{-\pare{2\Gamma+R}t_{ij}}
    \sum\limits_{\alpha=0}^\infty \pare{2\Gamma t_{ij}}^\alpha\frac{1}{\alpha!}  \pare{I_{\alpha-1/2}\pare{Rt_{ij}}-I_{\alpha+1/2}\pare{Rt_{ij}}},
\end{split}\ee
where $I_{\nu}\pare{z}$ is the modified Bessel function of the first kind.
Using Eq.~(4) (or Eq.~(6) which is the specific form we have) in Sec.~5.8.3 (Page 694) of \cite{Prudnikov},
\be\label{eq:BesselISeries}
    \sum\limits_{k=0}^\infty \frac{t^k}{k!}I_{\nu\pm k}\pare{z} = \pare{\frac{2t}{z}+1}^{\mp \nu/2}I_\nu\pare{\sqrt{z^2+2tz}}.
\ee
Plugging this into \eqnref{eq:X},
\be\begin{split}
    X\pare{t_i,t_j} 
    &= \sqrt{\frac{\pi Rt_{ij}}{2}}e^{-\pare{2\Gamma+R}t_{ij}} \pare{\pare{\frac{4\Gamma+R}{R}}^{1/4}I_{-1/2}\pare{t_{ij}\sqrt{R\pare{4\Gamma + R}}}-\pare{\frac{R}{4\Gamma+R}}^{1/4}I_{1/2}\pare{t_{ij}\sqrt{R\pare{4\Gamma + R}}}}\\
    &= 
    e^{-\pare{2\Gamma+R}t_{ij}}\pare{\cosh\pare{t_{ij}\sqrt{R\pare{4\Gamma + R}}}
    -\sqrt{\frac{R}{4\Gamma+R}}\sinh\pare{t_{ij}\sqrt{R\pare{4\Gamma + R}}}}.
\end{split}\ee

Now, going back to Eq.~\eqref{eq:P0_N2},
\be\begin{split}
    p_0\pare{t} = \sum\limits_{\beta=0}^\infty
    \pare{4R\Gamma}^\beta
    e^{-2Rt} * \left[ \pare{X*e^{-2R t}} *  \pare{X*e^{-2R t}} * ... \pare{\text{$\beta$ times}} \right].
\end{split}\ee
Performing a Laplace transform and using the property
\be\label{eq:LaplaceTransform_1_repeated}
    \mathcal{L}\left[f_1\pare{t}*f_2\pare{t}\right] = \mathcal{L}\left[f_1\pare{t}\right]\mathcal{L}\left[f_2\pare{t}\right] \equiv F_1\pare{s}F_2\pare{s},
\ee
where $F_j\pare{s}$ is the Laplace transform of $f_j\pare{t}$ gives
\be
    \mathcal{L}\left[X\pare{t}\right] = \frac{2\Gamma+s}{2Rs+\pare{s+2\Gamma}^2},
\ee
\be
    \mathcal{L}\left[e^{-2R t}\right] = \frac{1}{2R+s},
\ee
\be\begin{split}\label{eq:P0s_N2}
    \mathcal{L}\left[p_0\pare{t}\right]\pare{s} &= \frac{1}{2R+s}\sum\limits_{\beta=0}^\infty
    \pare{4R\Gamma}^\beta
    \pare{\frac{\pare{2\Gamma+s}}{\pare{2R+s}\pare{2Rs+\pare{s+2\Gamma}^2}}}^{\beta} \\
    &= \frac{2Rs+\pare{s+2\Gamma}^2}{s\pare{4R^2+4R\pare{s+\Gamma}+\pare{s+2\Gamma}^2}} \\
    &= \frac{\frac{\Gamma^2}{R^2+R\Gamma+\Gamma^2}}{s} + \frac{\frac{R}{2\pare{R-\sqrt{R\Gamma}+\Gamma}}}{s+2\pare{R-\sqrt{R\Gamma}+\Gamma}} + \frac{\frac{R}{2\pare{R+\sqrt{R\Gamma}+\Gamma}}}{s+2\pare{R+\sqrt{R\Gamma}+\Gamma}}.
\end{split}\ee
Taking an inverse Laplace transform, we get
\be
    p_0\pare{t} = \frac{\Gamma^2}{R^2+R\Gamma+\Gamma^2} + \frac{R}{2\pare{R-\sqrt{R\Gamma}+\Gamma}}e^{-2\pare{R-\sqrt{R\Gamma}+\Gamma}t} + \frac{R}{2\pare{R+\sqrt{R\Gamma}+\Gamma}}e^{-2\pare{R+\sqrt{R\Gamma}+\Gamma}t}.
\ee
We can use a very similar argument to solve for $p_1\pare{t}$ and $p_2\pare{t}$.
\be\begin{split}\label{eq:P1_N2}
    p_1\pare{t} &\!=\! 2R\sum\limits_{\beta=0}^\infty
    \pare{4R\Gamma}^\beta
    \!\!\!\!\!\!\!\!\!\!\!\!\!\!\!\!\!\!\int\limits_{0<t_1<...<t_{2\beta+1}<t} \!\!\!\!\!\!\!\!\!\!\!\!\!\!\!\!\! dt_1 ... dt_{2\beta+1} 
    X\pare{t_{2\beta}+1,t}
    e^{-2R\pare{t_{2\beta+1}-t_{2\beta}}}
    X\pare{t_{2\beta-1},t_{2\beta}}
    ...
    X\pare{t_3,t_4}
    e^{-2R\pare{t_3-t_2}}
    X\pare{t_1,t_2}
    e^{-2Rt_1} \\
    &= 2R\sum\limits_{\beta=0}^\infty
    \pare{4R\Gamma}^\beta
    \left[ \pare{X*e^{-2R t}} *  \pare{X*e^{-2R t}} * ... \pare{\text{$\beta+1$ times}} \right],
\end{split}\ee
\be\begin{split}
    \mathcal{L}\left[p_1\pare{t}\right]\pare{s} &= 2R\pare{\frac{\pare{2\Gamma+s}}{\pare{2R+s}\pare{2Rs+\pare{s+2\Gamma}^2}}}\sum\limits_{\beta=0}^\infty
    \pare{4R\Gamma}^\beta
    \pare{\frac{\pare{2\Gamma+s}}{\pare{2R+s}\pare{2Rs+\pare{s+2\Gamma}^2}}}^{\beta} \\
    &= \frac{2R\pare{s+2\Gamma}}{s\pare{4R^2+4R\pare{s+\Gamma}+\pare{s+2\Gamma}^2}}.
\end{split}\ee
\be\begin{split}
    \implies p_1\pare{t} &= \frac{R\Gamma}{R^2+R\Gamma+\Gamma^2} + \frac{R\pare{\sqrt{R}-\sqrt{\Gamma}}}{2\sqrt{\Gamma}\pare{R-\sqrt{R\Gamma}+\Gamma}} e^{-2\pare{R-\sqrt{R\Gamma}+\Gamma}t} - \frac{R\pare{\sqrt{R}+\sqrt{\Gamma}}}{2\sqrt{\Gamma}\pare{R+\sqrt{R\Gamma}+\Gamma}}e^{-2\pare{R+\sqrt{R\Gamma}+\Gamma}t}.
\end{split}\ee
\be\begin{split}\label{eq:P2_N2_integral}
    p_2\pare{t} &= 2R\sum\limits_{\beta=0}^\infty
    \pare{4R\Gamma}^\beta
   \!\!\!\!\!\!\!\!\!\!\!\!\! \!\!\!\!\int\limits_{0<t_1<...<t_{2\beta+1}<t} \!\!\!\!\!\!\!\!\!\!\!\!\!\!\! dt_1 ... dt_{2\beta+1} 
    \tilde{X}\pare{t_{2\beta+1},t}
    e^{-2R\pare{t_{2\beta+1}-t_{2\beta}}}
    X\pare{t_{2\beta-1},t_{2\beta}}
    ...
    X\pare{t_3,t_4}
    e^{-2R\pare{t_3-t_2}}
    X\pare{t_1,t_2}
    e^{-2Rt_1} \\
    &= 2R\sum\limits_{\beta=0}^\infty
    \pare{4R\Gamma}^\beta
    \tilde{X}\pare{t}*
    e^{-2Rt}*
    \left[ \pare{X*e^{-2R t}} *  \pare{X*e^{-2R t}} * ... \pare{\text{$\beta$ times}} \right],
\end{split}\ee
where 
\be\begin{split}
    \tilde{X}\pare{t_i,t_j} &= 2R\sum\limits_{\alpha=0}^\infty \pare{4R\Gamma}^\alpha 
    \left[ \pare{e^{-2\Gamma t_{ij}}*e^{-2\pare{\Gamma+R}t_{ij}}} * \pare{e^{-2\Gamma t_{ij}}*e^{-2\pare{\Gamma+R}t_{ij}}} * .... \pare{\text{$\alpha+1$ times}}\right]\\
    &= \pare{2R} e^{-2\Gamma t_{ij}}*X\pare{t_i,t_j}.
\end{split}\ee
\be\begin{split}\label{eq:P2_N2}
    \implies p_2\pare{t} &\!=\! 4R^2\sum\limits_{\beta=0}^\infty
    \pare{4R\Gamma}^\beta
    \int\limits_{0<t_1<...<t_{2\beta+1}<t} dt_1 ... dt_{2\beta+1}  
    \Bigg[e^{-2\Gamma\pare{t-t_{2\beta+1}}} * \\
    &X\pare{t_{2\beta+1},t}
    e^{-2R\pare{t_{2\beta+1}-t_{2\beta}}}
    X\pare{t_{2\beta-1},t_{2\beta}}
    ...
    X\pare{t_3,t_4}
    e^{-2R\pare{t_3-t_2}}
    X\pare{t_1,t_2}
    e^{-2Rt_1} \Bigg] \\
    &= 4R^2\sum\limits_{\beta=0}^\infty
    \pare{4R\Gamma}^\beta
    e^{-2\Gamma t}*
    \left[ \pare{X*e^{-2R t}} *  \pare{X*e^{-2R t}} * ... \pare{\text{$\beta+1$ times}} \right].
\end{split}\ee
\be\begin{split}
    \mathcal{L}\left[p_2\pare{t}\right]\pare{s} &= 4R^2\pare{\frac{\pare{2\Gamma+s}}{\pare{2\Gamma+s}\pare{2R+s}\pare{2Rs+\pare{s+2\Gamma}^2}}}\sum\limits_{\beta=0}^\infty
    \pare{4R\Gamma}^\beta
    \pare{\frac{\pare{2\Gamma+s}}{\pare{2R+s}\pare{2Rs+\pare{s+2\Gamma}^2}}}^{\beta} \\
    &= \frac{4R^2}{s\pare{4R^2+4R\pare{s+\Gamma}+\pare{s+2\Gamma}^2}}.
\end{split}\ee
\be\begin{split}
    \implies p_2\pare{t} &= \frac{R^2}{R^2+R\Gamma+\Gamma^2} - \frac{R^{3/2}}{2\sqrt{\Gamma}\pare{R-\sqrt{R\Gamma}+\Gamma}} e^{-2\pare{R-\sqrt{R\Gamma}+\Gamma}t} + \frac{R^{3/2}}{2\sqrt{\Gamma}\pare{R+\sqrt{R\Gamma}+\Gamma}}e^{-2\pare{R+\sqrt{R\Gamma}+\Gamma}t}.
\end{split}\ee

\subsection{Generalization to an arbitrary initial Dicke state}\label{app:anyinitalm}
For any $N$, starting in some initial state $\ket{m_0}$, we get an additional factor given by $\Gamma^{m_0} h_1...h_{m_0} Y_1\cdot Y_2...Y_{m_0}$, where
\be
    Y_n\pare{s} = U_n^2\pare{s}\frac{1}{1-\frac{C_{n+1}\pare{s}}{1-...\frac{C_{N-1}\pare{s}}{1-C_N\pare{s}}}}.
\ee
All trajectories that go from $\ket{m_0}\to\ket{0}\to\ket{k}$ would then be given by
\be\begin{split}
    P_{k,m_0}^{(0)}\pare{s} \!=\! \pare{h_1...h_k} \pare{h_1...h_{m_0}} R^k\Gamma^{m_0}\mathbf{U}_0\!\pare{s} \!\! 
    \spare{\frac{1}{1-\frac{C_{2}\pare{s}}{1-...\frac{C_{N-1}\pare{s}}{1-C_N\pare{s}}}}...\frac{1}{1-\frac{C_{k+1}\pare{s}}{1-...\frac{C_{N-1}\pare{s}}{1-C_N\pare{s}}}}}\!\!\!
    \spare{\frac{1}{1-\frac{C_{1}\pare{s}}{1-...\frac{C_{N-1}\pare{s}}{1-C_N\pare{s}}}}...\frac{1}{1-\frac{C_{m_0+1}\pare{s}}{1-...\frac{C_{N-1}\pare{s}}{1-C_N\pare{s}}}}}\!,
\end{split}\ee
where $\mathbf{U}_0\pare{s} = U_0^2\pare{s}U_1^2\pare{s}...U_k^2\pare{s} U_1^2\pare{s}...U_{m_0}^2\pare{s}$.
This can be rewritten in terms of a ratio of determinants:
\be\begin{split}
    P_{k,m_0}^{(0)}\pare{s}= \frac{\pare{N!}^2m_0!k!}{\pare{N-m_0}!\pare{N-k}!} R^k\Gamma^{m_0}\mathbf{U}_0\pare{s} 
    \frac{D_{k+2}\pare{s}D_{m_0+2}\pare{s}}{D_1\pare{s} D_2\pare{s}}.
\end{split}\ee
However, to account for all possible trajectories, we also need to add those that go from $\ket{m_0}\to\ket{a}\to\ket{k}$, $\forall 0<a\leq \min\pare{m_0,k}$, where $\ket{a}$ denotes the lowest state of the Dicke ladder visited in a given trajectory. For a given $a$, there are two distinct pieces: the first is a trajectory from $\ket{m_0}$ that reaches $\ket{a}$ for the first time and the second is one that goes from $\ket{a}$ to $\ket{k}$, never visiting any state below $\ket{a}$. All such paths are given by
\be\begin{split}
    P_{k,m_0}^{(a)}\!\!\pare{s} \!=\! \pare{h_{a+1}
    ...h_k} \pare{h_{a+1}...h_{m_0}}R^{k-a}\Gamma^{m_0-a}\mathbf{U}_a\!\pare{s} \!\! 
    \spare{\prod_{x=a+2}^{k+1}\frac{1}{1-\frac{C_{x}\pare{s}}{1-...\frac{C_{N-1}\pare{s}}{1-C_N\pare{s}}}}}\!\!\!
    \spare{\prod_{y=a+1}^{m_0+1}\frac{1}{1-\frac{C_{y}\pare{s}}{1-...\frac{C_{N-1}\pare{s}}{1-C_N\pare{s}}}}}\!,
\end{split}\ee
where $\mathbf{U}_a\pare{s}=U_a^2\pare{s}U_{a+1}^2\pare{s}...U_k^2\pare{s} U_{a+1}^2\pare{s}...U_{m_0}^2\pare{s}$. 

Thus, the Laplace-transformed population of state $\ket{k}$, starting in state $\ket{m_0}$ is
\be
    P_{k,m_0}\pare{s} = \sum\limits_{a=0}^{\min\pare{m_0,k}}P_{k,m_0}^{(a)}\!\!\pare{s} = \sum\limits_{a=0}^{\min\pare{m_0,k}} \pare{h_{a+1}
    ...h_k} \pare{h_{a+1}...h_{m_0}}R^{k-a}\Gamma^{m_0-a}\mathbf{U}_a\!\pare{s}\frac{D_{k+2}\pare{s}D_{m_0+2}\pare{s}}{D_{a+1}\pare{s} D_{a+2}\pare{s}}.
\ee
Note that $a$ is the minimum excitation number visited along the entire trajectory. Thus, summing over $a$ gives a partition of the possible histories without double counting.

\section{Time-dependent populations in the strong pumping regime}\label{app:onlyRsol}

As mentioned in the main text, in the strong pumping regime, the exact trajectory solution, to zeroth order in $\Gamma/R$, reduces to 
\begin{equation}
P_k^{(0)}(s)=
R^k\prod_{m=1}^{k}h_m
\prod_{j=1}^{k+1}
\frac{1}{s+Rh_j},
\quad 0\leq k<N,
\end{equation}
where we used
\begin{equation}
U_m^2(s)=
\frac{1}{s+Rh_{m+1}}
+\mathcal O\pare{\frac{\Gamma}{R^2}}.
\end{equation}
The inverse Laplace transform $p_k^{(0)}(t)$ can simply be obtained by recognizing the pole structure of the above expression: for even (odd) $N$, the poles are simple for $k<\bar{m}$, where $\bar{m} =N/2 \pare{(N+1)/2}$, while the solution also contains double poles otherwise. Using a partial fraction decomposition, the time-dependent populations can then be obtained. For even (odd) $N$ and $k<\bar{m}$, this is simply 
\begin{equation}
p_k^{(0)}(t)
=
\prod_{m=1}^{k}h_m
\sum_{j=1}^{k+1}
\frac{e^{-Rh_jt}}
{\displaystyle
\prod_{\substack{\ell=1\\ \ell\neq j}}^{k+1}
(h_\ell-h_j)}.
\label{eq:pure_pump_population}
\end{equation}
For $k\geq\bar{m}$, assuming $N$ is even,
\be
    p_k^{(0)}\pare{t} = \prod_{m=1}^{k}h_m\spare{\sum\limits_{j=1}^{N-k-1}A_je^{-Rh_jt}+\sum\limits_{j=N-k}^{\bar{m}}A_j\pare{Rt-B_j}e^{-Rh_jt}},
\ee
where $A_j = 1/\spare{\prod_{\substack{\ell=1\\ \ell\neq j}}^{N-k-1}(h_\ell-h_j)\prod_{\substack{\ell=N-k\\ \ell\neq j}}^{\bar{m}}(h_\ell-h_j)^2}$ and
$B_j=\sum_{q=1}^{N-k-1}1/\pare{h_q-h_j}+2\sum_{\substack{q=N-k\\ q\neq j}}^{\bar{m}}1/\pare{h_q-h_j}$. When $N$ is odd, the central pole $h_{\bar{m}}$ is simple, resulting in the population for $k\geq\bar{m}$ is given by 
\be
    p_k^{(0)}\pare{t} = \prod_{m=1}^{k}h_m\spare{\sum\limits_{j=1}^{N-k-1}{A_j}\pare{h_{\bar{m}}-h_j}e^{-Rh_jt}+A_{\bar{m}}e^{-Rh_{\bar{m}}t}+\sum\limits_{j=N-k}^{\bar{m}-1}{A_j}\pare{h_{\bar{m}}-h_j}\pare{Rt-{B_j}+\frac{1}{h_{\bar{m}}-h_j}}e^{-Rh_jt}}.
\ee
Finally, the population $p_N^{(0)}(t)$ is given by
\be
    p_N^{(0)}\pare{t} = 1-\sum\limits_{k=0}^{N-1}p_k^{(0)}\pare{t}.
\ee


\begin{thebibliography}{34}%
\makeatletter
\providecommand \@ifxundefined [1]{%
 \@ifx{#1\undefined}
}%
\providecommand \@ifnum [1]{%
 \ifnum #1\expandafter \@firstoftwo
 \else \expandafter \@secondoftwo
 \fi
}%
\providecommand \@ifx [1]{%
 \ifx #1\expandafter \@firstoftwo
 \else \expandafter \@secondoftwo
 \fi
}%
\providecommand \natexlab [1]{#1}%
\providecommand \enquote  [1]{``#1''}%
\providecommand \bibnamefont  [1]{#1}%
\providecommand \bibfnamefont [1]{#1}%
\providecommand \citenamefont [1]{#1}%
\providecommand \href@noop [0]{\@secondoftwo}%
\providecommand \href [0]{\begingroup \@sanitize@url \@href}%
\providecommand \@href[1]{\@@startlink{#1}\@@href}%
\providecommand \@@href[1]{\endgroup#1\@@endlink}%
\providecommand \@sanitize@url [0]{\catcode `\\12\catcode `\$12\catcode `\&12\catcode `\#12\catcode `\^12\catcode `\_12\catcode `\%12\relax}%
\providecommand \@@startlink[1]{}%
\providecommand \@@endlink[0]{}%
\providecommand \url  [0]{\begingroup\@sanitize@url \@url }%
\providecommand \@url [1]{\endgroup\@href {#1}{\urlprefix }}%
\providecommand \urlprefix  [0]{URL }%
\providecommand \Eprint [0]{\href }%
\providecommand \doibase [0]{http://dx.doi.org/}%
\providecommand \selectlanguage [0]{\@gobble}%
\providecommand \bibinfo  [0]{\@secondoftwo}%
\providecommand \bibfield  [0]{\@secondoftwo}%
\providecommand \translation [1]{[#1]}%
\providecommand \BibitemOpen [0]{}%
\providecommand \bibitemStop [0]{}%
\providecommand \bibitemNoStop [0]{.\EOS\space}%
\providecommand \EOS [0]{\spacefactor3000\relax}%
\providecommand \BibitemShut  [1]{\csname bibitem#1\endcsname}%
\let\auto@bib@innerbib\@empty
\bibitem [{\citenamefont {Dicke}(1954)}]{Dicke1954}%
  \BibitemOpen
  \bibfield  {author} {\bibinfo {author} {\bibfnamefont {R.~H.}\ \bibnamefont {Dicke}},\ }\href {\doibase 10.1103/PhysRev.93.99} {\bibfield  {journal} {\bibinfo  {journal} {Phys. Rev.}\ }\textbf {\bibinfo {volume} {93}},\ \bibinfo {pages} {99} (\bibinfo {year} {1954})}\BibitemShut {NoStop}%
\bibitem [{\citenamefont {Rehler}\ and\ \citenamefont {Eberly}(1971)}]{RehlerEberly1971}%
  \BibitemOpen
  \bibfield  {author} {\bibinfo {author} {\bibfnamefont {N.~E.}\ \bibnamefont {Rehler}}\ and\ \bibinfo {author} {\bibfnamefont {J.~H.}\ \bibnamefont {Eberly}},\ }\href {\doibase 10.1103/PhysRevA.3.1735} {\bibfield  {journal} {\bibinfo  {journal} {Phys. Rev. A}\ }\textbf {\bibinfo {volume} {3}},\ \bibinfo {pages} {1735} (\bibinfo {year} {1971})}\BibitemShut {NoStop}%
\bibitem [{\citenamefont {Bonifacio}\ and\ \citenamefont {Lugiato}(1975)}]{BonifacioLugiato1975}%
  \BibitemOpen
  \bibfield  {author} {\bibinfo {author} {\bibfnamefont {R.}~\bibnamefont {Bonifacio}}\ and\ \bibinfo {author} {\bibfnamefont {L.~A.}\ \bibnamefont {Lugiato}},\ }\href {\doibase 10.1103/PhysRevA.11.1507} {\bibfield  {journal} {\bibinfo  {journal} {Phys. Rev. A}\ }\textbf {\bibinfo {volume} {11}},\ \bibinfo {pages} {1507} (\bibinfo {year} {1975})}\BibitemShut {NoStop}%
\bibitem [{\citenamefont {Gross}\ and\ \citenamefont {Haroche}(1982)}]{GrossHaroche1982}%
  \BibitemOpen
  \bibfield  {author} {\bibinfo {author} {\bibfnamefont {M.}~\bibnamefont {Gross}}\ and\ \bibinfo {author} {\bibfnamefont {S.}~\bibnamefont {Haroche}},\ }\href {\doibase https://doi.org/10.1016/0370-1573(82)90102-8} {\bibfield  {journal} {\bibinfo  {journal} {Physics Reports}\ }\textbf {\bibinfo {volume} {93}},\ \bibinfo {pages} {301} (\bibinfo {year} {1982})}\BibitemShut {NoStop}%
\bibitem [{\citenamefont {Shammah}\ \emph {et~al.}(2018)\citenamefont {Shammah}, \citenamefont {Ahmed}, \citenamefont {Lambert}, \citenamefont {De~Liberato},\ and\ \citenamefont {Nori}}]{ShammahEtAl2018}%
  \BibitemOpen
  \bibfield  {author} {\bibinfo {author} {\bibfnamefont {N.}~\bibnamefont {Shammah}}, \bibinfo {author} {\bibfnamefont {S.}~\bibnamefont {Ahmed}}, \bibinfo {author} {\bibfnamefont {N.}~\bibnamefont {Lambert}}, \bibinfo {author} {\bibfnamefont {S.}~\bibnamefont {De~Liberato}}, \ and\ \bibinfo {author} {\bibfnamefont {F.}~\bibnamefont {Nori}},\ }\href {\doibase 10.1103/PhysRevA.98.063815} {\bibfield  {journal} {\bibinfo  {journal} {Phys. Rev. A}\ }\textbf {\bibinfo {volume} {98}},\ \bibinfo {pages} {063815} (\bibinfo {year} {2018})}\BibitemShut {NoStop}%
\bibitem [{\citenamefont {Holzinger}\ \emph {et~al.}(2026{\natexlab{a}})\citenamefont {Holzinger}, \citenamefont {Bassler}, \citenamefont {Lyne}, \citenamefont {Jimenez}, \citenamefont {Gohsrich},\ and\ \citenamefont {Genes}}]{HolzingerEtAl2025}%
  \BibitemOpen
  \bibfield  {author} {\bibinfo {author} {\bibfnamefont {R.}~\bibnamefont {Holzinger}}, \bibinfo {author} {\bibfnamefont {N.~S.}\ \bibnamefont {Bassler}}, \bibinfo {author} {\bibfnamefont {J.}~\bibnamefont {Lyne}}, \bibinfo {author} {\bibfnamefont {F.~G.}\ \bibnamefont {Jimenez}}, \bibinfo {author} {\bibfnamefont {J.~T.}\ \bibnamefont {Gohsrich}}, \ and\ \bibinfo {author} {\bibfnamefont {C.}~\bibnamefont {Genes}},\ }\href {\doibase 10.1103/9p44-kwfl} {\bibfield  {journal} {\bibinfo  {journal} {Phys. Rev. A}\ }\textbf {\bibinfo {volume} {113}},\ \bibinfo {pages} {043718} (\bibinfo {year} {2026}{\natexlab{a}})}\BibitemShut {NoStop}%
\bibitem [{\citenamefont {Holzinger}\ and\ \citenamefont {Genes}(2025)}]{HolzingerGenes2024}%
  \BibitemOpen
  \bibfield  {author} {\bibinfo {author} {\bibfnamefont {R.}~\bibnamefont {Holzinger}}\ and\ \bibinfo {author} {\bibfnamefont {C.}~\bibnamefont {Genes}},\ }\href {\doibase 10.1515/zna-2025-0179} {\bibfield  {journal} {\bibinfo  {journal} {Z. Naturforsch. A}\ }\textbf {\bibinfo {volume} {80}},\ \bibinfo {pages} {673} (\bibinfo {year} {2025})}\BibitemShut {NoStop}%
\bibitem [{\citenamefont {Holzinger}\ \emph {et~al.}(2026{\natexlab{b}})\citenamefont {Holzinger}, \citenamefont {Bassler}, \citenamefont {Lyne}, \citenamefont {Yelin},\ and\ \citenamefont {Genes}}]{holzinger2026symbolic}%
  \BibitemOpen
  \bibfield  {author} {\bibinfo {author} {\bibfnamefont {R.}~\bibnamefont {Holzinger}}, \bibinfo {author} {\bibfnamefont {N.~S.}\ \bibnamefont {Bassler}}, \bibinfo {author} {\bibfnamefont {J.}~\bibnamefont {Lyne}}, \bibinfo {author} {\bibfnamefont {S.~F.}\ \bibnamefont {Yelin}}, \ and\ \bibinfo {author} {\bibfnamefont {C.}~\bibnamefont {Genes}},\ }\href {https://arxiv.org/abs/2511.02390} {\enquote {\bibinfo {title} {Symbolic quantum-trajectory method for multichannel dicke superradiance},}\ } (\bibinfo {year} {2026}{\natexlab{b}}),\ \Eprint {http://arxiv.org/abs/2511.02390} {arXiv:2511.02390 [quant-ph]} \BibitemShut {NoStop}%
\bibitem [{\citenamefont {Lee}(1977{\natexlab{a}})}]{Lee1977I}%
  \BibitemOpen
  \bibfield  {author} {\bibinfo {author} {\bibfnamefont {C.~T.}\ \bibnamefont {Lee}},\ }\href {\doibase 10.1103/PhysRevA.15.2019} {\bibfield  {journal} {\bibinfo  {journal} {Phys. Rev. A}\ }\textbf {\bibinfo {volume} {15}},\ \bibinfo {pages} {2019} (\bibinfo {year} {1977}{\natexlab{a}})}\BibitemShut {NoStop}%
\bibitem [{\citenamefont {Lee}(1977{\natexlab{b}})}]{Lee1977II}%
  \BibitemOpen
  \bibfield  {author} {\bibinfo {author} {\bibfnamefont {C.~T.}\ \bibnamefont {Lee}},\ }\href {\doibase 10.1103/PhysRevA.16.301} {\bibfield  {journal} {\bibinfo  {journal} {Phys. Rev. A}\ }\textbf {\bibinfo {volume} {16}},\ \bibinfo {pages} {301} (\bibinfo {year} {1977}{\natexlab{b}})}\BibitemShut {NoStop}%
\bibitem [{\citenamefont {Malz}\ \emph {et~al.}(2022)\citenamefont {Malz}, \citenamefont {Trivedi},\ and\ \citenamefont {Cirac}}]{MalzTrivediCirac2022}%
  \BibitemOpen
  \bibfield  {author} {\bibinfo {author} {\bibfnamefont {D.}~\bibnamefont {Malz}}, \bibinfo {author} {\bibfnamefont {R.}~\bibnamefont {Trivedi}}, \ and\ \bibinfo {author} {\bibfnamefont {J.~I.}\ \bibnamefont {Cirac}},\ }\href {\doibase 10.1103/PhysRevA.106.013716} {\bibfield  {journal} {\bibinfo  {journal} {Phys. Rev. A}\ }\textbf {\bibinfo {volume} {106}},\ \bibinfo {pages} {013716} (\bibinfo {year} {2022})}\BibitemShut {NoStop}%
\bibitem [{\citenamefont {Puri}\ and\ \citenamefont {Lawande}(1979)}]{puri1979exact}%
  \BibitemOpen
  \bibfield  {author} {\bibinfo {author} {\bibfnamefont {R.}~\bibnamefont {Puri}}\ and\ \bibinfo {author} {\bibfnamefont {S.}~\bibnamefont {Lawande}},\ }\href@noop {} {\bibfield  {journal} {\bibinfo  {journal} {Physics Letters A}\ }\textbf {\bibinfo {volume} {72}},\ \bibinfo {pages} {200} (\bibinfo {year} {1979})}\BibitemShut {NoStop}%
\bibitem [{\citenamefont {Kilin}(1980)}]{kilin1980cooperative}%
  \BibitemOpen
  \bibfield  {author} {\bibinfo {author} {\bibfnamefont {S.~J.}\ \bibnamefont {Kilin}},\ }\href@noop {} {\bibfield  {journal} {\bibinfo  {journal} {Journal of Physics B: Atomic and Molecular Physics}\ }\textbf {\bibinfo {volume} {13}},\ \bibinfo {pages} {2653} (\bibinfo {year} {1980})}\BibitemShut {NoStop}%
\bibitem [{\citenamefont {Drummond}(1980)}]{drummond1980observables}%
  \BibitemOpen
  \bibfield  {author} {\bibinfo {author} {\bibfnamefont {P.~D.}\ \bibnamefont {Drummond}},\ }\href {\doibase 10.1103/PhysRevA.22.1179} {\bibfield  {journal} {\bibinfo  {journal} {Phys. Rev. A}\ }\textbf {\bibinfo {volume} {22}},\ \bibinfo {pages} {1179} (\bibinfo {year} {1980})}\BibitemShut {NoStop}%
\bibitem [{\citenamefont {Ribeiro}\ and\ \citenamefont {Prosen}(2019)}]{RibeiroProsen2019}%
  \BibitemOpen
  \bibfield  {author} {\bibinfo {author} {\bibfnamefont {P.}~\bibnamefont {Ribeiro}}\ and\ \bibinfo {author} {\bibfnamefont {T.}~\bibnamefont {Prosen}},\ }\href {\doibase 10.1103/PhysRevLett.122.010401} {\bibfield  {journal} {\bibinfo  {journal} {Physical Review Letters}\ }\textbf {\bibinfo {volume} {122}},\ \bibinfo {pages} {010401} (\bibinfo {year} {2019})},\ \Eprint {http://arxiv.org/abs/1807.09438} {arXiv:1807.09438 [quant-ph]} \BibitemShut {NoStop}%
\bibitem [{\citenamefont {Kubo}(1962)}]{Kubo1962}%
  \BibitemOpen
  \bibfield  {author} {\bibinfo {author} {\bibfnamefont {R.}~\bibnamefont {Kubo}},\ }\href {\doibase 10.1143/JPSJ.17.1100} {\bibfield  {journal} {\bibinfo  {journal} {Journal of the Physical Society of Japan}\ }\textbf {\bibinfo {volume} {17}},\ \bibinfo {pages} {1100} (\bibinfo {year} {1962})}\BibitemShut {NoStop}%
\bibitem [{\citenamefont {Bonifacio}\ and\ \citenamefont {Lugiato}(1978)}]{BonifacioLugiato1978}%
  \BibitemOpen
  \bibfield  {author} {\bibinfo {author} {\bibfnamefont {R.}~\bibnamefont {Bonifacio}}\ and\ \bibinfo {author} {\bibfnamefont {L.~A.}\ \bibnamefont {Lugiato}},\ }\href {\doibase 10.1103/PhysRevA.18.1129} {\bibfield  {journal} {\bibinfo  {journal} {Phys. Rev. A}\ }\textbf {\bibinfo {volume} {18}},\ \bibinfo {pages} {1129} (\bibinfo {year} {1978})}\BibitemShut {NoStop}%
\bibitem [{\citenamefont {Kr{\"a}mer}\ and\ \citenamefont {Ritsch}(2015)}]{KramerRitsch2015}%
  \BibitemOpen
  \bibfield  {author} {\bibinfo {author} {\bibfnamefont {S.}~\bibnamefont {Kr{\"a}mer}}\ and\ \bibinfo {author} {\bibfnamefont {H.}~\bibnamefont {Ritsch}},\ }\href {\doibase 10.1140/epjd/e2015-60266-5} {\bibfield  {journal} {\bibinfo  {journal} {Eur. Phys. J. D}\ }\textbf {\bibinfo {volume} {69}},\ \bibinfo {pages} {282} (\bibinfo {year} {2015})}\BibitemShut {NoStop}%
\bibitem [{\citenamefont {Barberena}\ and\ \citenamefont {Rey}(2024)}]{BarberenaRey2024}%
  \BibitemOpen
  \bibfield  {author} {\bibinfo {author} {\bibfnamefont {D.}~\bibnamefont {Barberena}}\ and\ \bibinfo {author} {\bibfnamefont {A.~M.}\ \bibnamefont {Rey}},\ }\href {\doibase 10.1103/PhysRevA.109.013709} {\bibfield  {journal} {\bibinfo  {journal} {Phys. Rev. A}\ }\textbf {\bibinfo {volume} {109}},\ \bibinfo {pages} {013709} (\bibinfo {year} {2024})}\BibitemShut {NoStop}%
\bibitem [{\citenamefont {Dalibard}\ \emph {et~al.}(1992)\citenamefont {Dalibard}, \citenamefont {Castin},\ and\ \citenamefont {M{\o}lmer}}]{DalibardCastinMolmer1992}%
  \BibitemOpen
  \bibfield  {author} {\bibinfo {author} {\bibfnamefont {J.}~\bibnamefont {Dalibard}}, \bibinfo {author} {\bibfnamefont {Y.}~\bibnamefont {Castin}}, \ and\ \bibinfo {author} {\bibfnamefont {K.}~\bibnamefont {M{\o}lmer}},\ }\href {\doibase 10.1103/PhysRevLett.68.580} {\bibfield  {journal} {\bibinfo  {journal} {Phys. Rev. Lett.}\ }\textbf {\bibinfo {volume} {68}},\ \bibinfo {pages} {580} (\bibinfo {year} {1992})}\BibitemShut {NoStop}%
\bibitem [{\citenamefont {Shimshi}\ and\ \citenamefont {Shahmoon}(2026)}]{ShimshiShahmoon2024}%
  \BibitemOpen
  \bibfield  {author} {\bibinfo {author} {\bibfnamefont {Y.}~\bibnamefont {Shimshi}}\ and\ \bibinfo {author} {\bibfnamefont {E.}~\bibnamefont {Shahmoon}},\ }\href {\doibase 10.1103/rp9s-tgp7} {\bibfield  {journal} {\bibinfo  {journal} {Phys. Rev. Lett.}\ }\textbf {\bibinfo {volume} {137}},\ \bibinfo {pages} {103601} (\bibinfo {year} {2026})}\BibitemShut {NoStop}%
\bibitem [{\citenamefont {Karlin}\ and\ \citenamefont {McGregor}(1957)}]{karlin1957classification}%
  \BibitemOpen
  \bibfield  {author} {\bibinfo {author} {\bibfnamefont {S.}~\bibnamefont {Karlin}}\ and\ \bibinfo {author} {\bibfnamefont {J.}~\bibnamefont {McGregor}},\ }\href {\doibase 10.1090/s0002-9947-1957-0094854-8} {\bibfield  {journal} {\bibinfo  {journal} {Transactions of the American Mathematical Society}\ }\textbf {\bibinfo {volume} {86}},\ \bibinfo {pages} {366–400} (\bibinfo {year} {1957})}\BibitemShut {NoStop}%
\bibitem [{\citenamefont {Karlin}\ and\ \citenamefont {McGregor}(1958)}]{karlin1958linear}%
  \BibitemOpen
  \bibfield  {author} {\bibinfo {author} {\bibfnamefont {S.}~\bibnamefont {Karlin}}\ and\ \bibinfo {author} {\bibfnamefont {J.}~\bibnamefont {McGregor}},\ }\href {http://www.jstor.org/stable/24900526} {\bibfield  {journal} {\bibinfo  {journal} {Journal of Mathematics and Mechanics}\ }\textbf {\bibinfo {volume} {7}},\ \bibinfo {pages} {643} (\bibinfo {year} {1958})}\BibitemShut {NoStop}%
\bibitem [{\citenamefont {Ismail}\ \emph {et~al.}(1990)\citenamefont {Ismail}, \citenamefont {Letessier}, \citenamefont {Masson},\ and\ \citenamefont {Valent}}]{ismail1990birth}%
  \BibitemOpen
  \bibfield  {author} {\bibinfo {author} {\bibfnamefont {M.~E.~H.}\ \bibnamefont {Ismail}}, \bibinfo {author} {\bibfnamefont {J.}~\bibnamefont {Letessier}}, \bibinfo {author} {\bibfnamefont {D.~R.}\ \bibnamefont {Masson}}, \ and\ \bibinfo {author} {\bibfnamefont {G.}~\bibnamefont {Valent}},\ }\enquote {\bibinfo {title} {Birth and death processes and orthogonal polynomials},}\ in\ \href {\doibase 10.1007/978-94-009-0501-6_11} {\emph {\bibinfo {booktitle} {Orthogonal Polynomials}}}\ (\bibinfo  {publisher} {Springer Netherlands},\ \bibinfo {year} {1990})\ p.\ \bibinfo {pages} {229–255}\BibitemShut {NoStop}%
\bibitem [{\citenamefont {Guillemin}\ and\ \citenamefont {Pinchon}(1999)}]{guillemin1999excursion}%
  \BibitemOpen
  \bibfield  {author} {\bibinfo {author} {\bibfnamefont {F.}~\bibnamefont {Guillemin}}\ and\ \bibinfo {author} {\bibfnamefont {D.}~\bibnamefont {Pinchon}},\ }\href {\doibase 10.1239/jap/1032374632} {\bibfield  {journal} {\bibinfo  {journal} {Journal of Applied Probability}\ }\textbf {\bibinfo {volume} {36}},\ \bibinfo {pages} {752–770} (\bibinfo {year} {1999})}\BibitemShut {NoStop}%
\bibitem [{\citenamefont {Sasaki}(2009)}]{sasaki2009exactly}%
  \BibitemOpen
  \bibfield  {author} {\bibinfo {author} {\bibfnamefont {R.}~\bibnamefont {Sasaki}},\ }\href {\doibase 10.1063/1.3215983} {\bibfield  {journal} {\bibinfo  {journal} {Journal of Mathematical Physics}\ }\textbf {\bibinfo {volume} {50}} (\bibinfo {year} {2009}),\ 10.1063/1.3215983}\BibitemShut {NoStop}%
\bibitem [{\citenamefont {Koekoek}\ and\ \citenamefont {Swarttouw}(1996)}]{koekoek1996askey}%
  \BibitemOpen
  \bibfield  {author} {\bibinfo {author} {\bibfnamefont {R.}~\bibnamefont {Koekoek}}\ and\ \bibinfo {author} {\bibfnamefont {R.~F.}\ \bibnamefont {Swarttouw}},\ }\href {\doibase 10.48550/ARXIV.MATH/9602214} {\enquote {\bibinfo {title} {The askey-scheme of hypergeometric orthogonal polynomials and its q-analogue},}\ } (\bibinfo {year} {1996})\BibitemShut {NoStop}%
\bibitem [{\citenamefont {Koekoek}\ \emph {et~al.}(2010)\citenamefont {Koekoek}, \citenamefont {Lesky},\ and\ \citenamefont {Swarttouw}}]{koekoek2010hypergeometric}%
  \BibitemOpen
  \bibfield  {author} {\bibinfo {author} {\bibfnamefont {R.}~\bibnamefont {Koekoek}}, \bibinfo {author} {\bibfnamefont {P.~A.}\ \bibnamefont {Lesky}}, \ and\ \bibinfo {author} {\bibfnamefont {R.~F.}\ \bibnamefont {Swarttouw}},\ }\enquote {\bibinfo {title} {Hypergeometric orthogonal polynomials},}\ in\ \href {\doibase 10.1007/978-3-642-05014-5_9} {\emph {\bibinfo {booktitle} {Hypergeometric Orthogonal Polynomials and Their q-Analogues}}}\ (\bibinfo  {publisher} {Springer Berlin Heidelberg},\ \bibinfo {year} {2010})\ p.\ \bibinfo {pages} {183–253}\BibitemShut {NoStop}%
\bibitem [{\citenamefont {Roberts}\ \emph {et~al.}(2021)\citenamefont {Roberts}, \citenamefont {Lingenfelter},\ and\ \citenamefont {Clerk}}]{RobertsLingenfelterClerk2021}%
  \BibitemOpen
  \bibfield  {author} {\bibinfo {author} {\bibfnamefont {D.}~\bibnamefont {Roberts}}, \bibinfo {author} {\bibfnamefont {A.}~\bibnamefont {Lingenfelter}}, \ and\ \bibinfo {author} {\bibfnamefont {A.~A.}\ \bibnamefont {Clerk}},\ }\href {\doibase 10.1103/PRXQuantum.2.020336} {\bibfield  {journal} {\bibinfo  {journal} {PRX Quantum}\ }\textbf {\bibinfo {volume} {2}},\ \bibinfo {pages} {020336} (\bibinfo {year} {2021})},\ \Eprint {http://arxiv.org/abs/2011.02148} {arXiv:2011.02148 [quant-ph]} \BibitemShut {NoStop}%
\bibitem [{\citenamefont {Roberts}\ and\ \citenamefont {Clerk}(2023{\natexlab{a}})}]{RobertsClerk2023Ising}%
  \BibitemOpen
  \bibfield  {author} {\bibinfo {author} {\bibfnamefont {D.}~\bibnamefont {Roberts}}\ and\ \bibinfo {author} {\bibfnamefont {A.~A.}\ \bibnamefont {Clerk}},\ }\href {\doibase 10.1103/PhysRevLett.131.190403} {\bibfield  {journal} {\bibinfo  {journal} {Physical Review Letters}\ }\textbf {\bibinfo {volume} {131}},\ \bibinfo {pages} {190403} (\bibinfo {year} {2023}{\natexlab{a}})},\ \Eprint {http://arxiv.org/abs/2307.06946} {arXiv:2307.06946 [quant-ph]} \BibitemShut {NoStop}%
\bibitem [{\citenamefont {Yao}\ \emph {et~al.}(2025)\citenamefont {Yao}, \citenamefont {Lingenfelter}, \citenamefont {Belyansky}, \citenamefont {Roberts},\ and\ \citenamefont {Clerk}}]{YaoEtAl2025}%
  \BibitemOpen
  \bibfield  {author} {\bibinfo {author} {\bibfnamefont {M.}~\bibnamefont {Yao}}, \bibinfo {author} {\bibfnamefont {A.}~\bibnamefont {Lingenfelter}}, \bibinfo {author} {\bibfnamefont {R.}~\bibnamefont {Belyansky}}, \bibinfo {author} {\bibfnamefont {D.}~\bibnamefont {Roberts}}, \ and\ \bibinfo {author} {\bibfnamefont {A.~A.}\ \bibnamefont {Clerk}},\ }\href {\doibase 10.1103/PhysRevLett.134.130404} {\bibfield  {journal} {\bibinfo  {journal} {Physical Review Letters}\ }\textbf {\bibinfo {volume} {134}},\ \bibinfo {pages} {130404} (\bibinfo {year} {2025})},\ \Eprint {http://arxiv.org/abs/2407.12750} {arXiv:2407.12750 [quant-ph]} \BibitemShut {NoStop}%
\bibitem [{\citenamefont {Hartmann}(2026)}]{hartmann2026exactdampingbasiscontinuedfractions}%
  \BibitemOpen
  \bibfield  {author} {\bibinfo {author} {\bibfnamefont {S.}~\bibnamefont {Hartmann}},\ }\href {https://arxiv.org/abs/2609.17785} {\enquote {\bibinfo {title} {Exact damping-basis continued fractions for few-atom lasers: collective correlations beyond second-order cumulant closure},}\ } (\bibinfo {year} {2026}),\ \Eprint {http://arxiv.org/abs/2609.17785} {arXiv:2609.17785 [quant-ph]} \BibitemShut {NoStop}%
\bibitem [{\citenamefont {Roberts}\ and\ \citenamefont {Clerk}(2023{\natexlab{b}})}]{RobertsClerk2023Bosonic}%
  \BibitemOpen
  \bibfield  {author} {\bibinfo {author} {\bibfnamefont {D.}~\bibnamefont {Roberts}}\ and\ \bibinfo {author} {\bibfnamefont {A.~A.}\ \bibnamefont {Clerk}},\ }\href {\doibase 10.1103/PhysRevLett.130.063601} {\bibfield  {journal} {\bibinfo  {journal} {Physical Review Letters}\ }\textbf {\bibinfo {volume} {130}},\ \bibinfo {pages} {063601} (\bibinfo {year} {2023}{\natexlab{b}})},\ \Eprint {http://arxiv.org/abs/2208.05451} {arXiv:2208.05451 [quant-ph]} \BibitemShut {NoStop}%
\bibitem [{\citenamefont {Prudnikov}\ \emph {et~al.}(1986)\citenamefont {Prudnikov}, \citenamefont {Brychkov},\ and\ \citenamefont {Marichev}}]{Prudnikov}%
  \BibitemOpen
  \bibfield  {author} {\bibinfo {author} {\bibfnamefont {A.~P.}\ \bibnamefont {Prudnikov}}, \bibinfo {author} {\bibfnamefont {I.~A.}\ \bibnamefont {Brychkov}}, \ and\ \bibinfo {author} {\bibfnamefont {O.~I.}\ \bibnamefont {Marichev}},\ }\href@noop {} {\emph {\bibinfo {title} {Integrals and series}}}\ (\bibinfo  {publisher} {Gordon and Breach Science Publishers},\ \bibinfo {address} {New York},\ \bibinfo {year} {1986})\BibitemShut {NoStop}%
\end{thebibliography}
\end{document}